\documentclass{PoS}

\usepackage{amsmath,amssymb}

\newcommand{\bea}{\begin{eqnarray}}
\newcommand{\beas}{\begin{eqnarray*}}

\newcommand{\eea}{\end{eqnarray}}
\newcommand{\eeas}{\end{eqnarray*}}
\newcommand{\bd}{\begin{displaymath}}
\newcommand{\ed}{\end{displaymath}}
\newcommand{\be}{\begin{equation}}
\newcommand{\ee}{\end{equation}}
\newcommand{\bi}{\begin{itemize}}
\newcommand{\ei}{\end{itemize}}

\newcommand{\ff}{f\hspace{-0.3cm}f}

\newcommand{\captionfonts}{\small}
\makeatletter  % Allow the use of @ in command names
\long\def\@makecaption#1#2{%
  \vskip\abovecaptionskip
  \sbox\@tempboxa{{\captionfonts #1: #2}}%
  \ifdim \wd\@tempboxa >\hsize
    {\captionfonts #1: #2\par}
  \else
    \hbox to\hsize{\hfil\box\@tempboxa\hfil}%
  \fi
  \vskip\belowcaptionskip}
\makeatother   % Cancel the effect of \makeatletter

\title{Soft gluon resummation for slepton pair-production}

\ShortTitle{Soft gluon resummation for slepton pair-production}

\author{\speaker{Alessandro Broggio}\\
        {\sl
Institut f\"ur Physik (THEP),\\ Johannes Gutenberg-Universit\"at, D--55099 Mainz, Germany}\\
        E-mail: \email{broggio@uni-mainz.de}}
        
\author{Matthias Neubert\\
        {\sl
Institut f\"ur Physik (THEP),\\ Johannes Gutenberg-Universit\"at, D--55099 Mainz, Germany}\\
        E-mail: \email{matthias.neubert@uni-mainz.de}}

\author{Leonardo Vernazza\footnote{Alexander-von-Humboldt Fellow}\\
        {\sl
Institut f\"ur Physik (THEP),\\ Johannes Gutenberg-Universit\"at, D--55099 Mainz, Germany}\\
        E-mail: \email{vernazza@uni-mainz.de}}

\abstract{We report on recent results on the differential 
cross section for slepton pair-production at hadron colliders. 
We use an approach to threshold resummation, based 
on soft-collinear effective theory, to quantify the dynamical
enhancement of the partonic threshold region. We evaluate the resummed invariant mass distribution 
and total cross section at next-to-next-to-next-to-leading 
logarithmic order, and match the result onto next-to-leading 
order calculation.}

\FullConference{The 2011 Europhysics Conference on High Energy Physics-HEP 2011,\\
		July 21-27, 2011\\
		Grenoble, Rh\^one-Alpes, France}

\begin{document}

\section{Introduction}
With the advent of the Large Hadron Collider (LHC),
the experimental investigation of TeV-scale physics
is now fully accessible. The stabilization
of the electroweak scale requires new particles and
interactions in the TeV range, and supersymmetry is
one of the most compelling scenario which achieves
such a stabilization. It predicts the existence of "supersymmetric
partners" for all of the SM particles. The superpartners are expected
to have masses of the order of a TeV.

%It introduces, beside every
%Standard Model (SM) particle, a corresponding
%superpartner which is expected to have mass
%in the TeV range.

The scalar-leptons are expected to be among the lightest 
supersymmetric particles and, in many scenarios, they 
decay directly into their corresponding SM partners and 
the lightest supersymmetric particle. The production of 
a slepton-pair should have quite simple signatures, for example, 
a couple of energetic leptons plus missing energy.

Currently, complete next-to-leading order (NLO) 
calculations are available in SUSY-QCD for the total 
slepton-pair production cross section \cite{Beenakker:1999xh}, 
as well for the differential distributions \cite{Bozzi:2007qr}. 
Reducing the scale uncertainty by computing higher order 
terms in the perturbative expansion is a task which quickly becomes 
challenging due to the number of different scales 
which enter the problem. A good alternative, to improve 
the theoretical prediction, is to resort to threshold 
resummation methods \cite{Mainz1}, which allows one to take into 
account the dominant contributions of the higher 
order terms. Expressions for the resummed invariant 
mass distribution and total cross section were obtained 
in Mellin moment space at next-to-leading logarithmic 
accuracy in \cite{Bozzi:2007qr}. We extend this result to 
next-to-next-to-next-to-leading logarithmic (NNNLL) 
accuracy using soft-collinear effective theory tecniques. 
In this work we summarize the results reported in 
\cite{Mainz1}.

\section{Differential cross section}

We consider the process
\be\label{process}
N_1(P_1)+N_2(P_2) \to \gamma^*,Z^{0*}+X \to \tilde l(p_3), \tilde l^*(p_4) + X\, ,
\ee
where $N_{1}, N_{2}$ indicate two protons (LHC case) 
and $X$ represents an inclusive hadronic final state. 
The doubly-differential cross section in the invariant 
mass $M^{2}$ and in the rapidity $Y$ for the process in 
Eq.~(\ref{process}) can be written as
\be\label{doublydiff}
\frac{d^{2}\sigma}{d M^{2}dY}=
\frac{\pi \alpha_{\text{em}} \beta^{3}_{\tilde{l}}}{3 N_{c} M^{2} s}
\sum_{i j}\int d x_{1} d x_{2} 
\tilde{C}_{i j}(x_{1},x_{2},\hat{s},M,m_{\tilde{q}},m_{\tilde{g}},\mu_{f})
f_{i/N_{1}}(x_{1},\mu_{f})f_{j/N_{2}}(x_{2},\mu_{f}) \, ,
\ee
where $f_{i/N}$ are the parton distribution functions, 
$\tilde{C}_{ij}$ are the hard-scattering kernels, $s$ 
$(\hat{s})$ is the hadronic (partonic) center of mass 
energy squared and 
$\beta_{\tilde{l}}=\sqrt{1-4 m^{2}_{\tilde{l}}/M^{2}}$. 
At NLO the sum extends over the following channels: 
$(ij) = (q \bar{q})$, $(\bar{q} q)$, $(q g)$, $(g q)$, 
$(\bar{q} g)$, $(g \bar{q})$. We are interested in the 
evaluation of the SUSY-QCD corrections near threshold, 
for this pourpose it is useful to introduce the 
following quantities:
\be
\tau = \frac{M^{2}}{s}\,, \quad 
z = \frac{M^{2}}{\hat{s}}=\frac{\tau}{x_{1} x_{2}}\, ,
\ee
where $\hat{s} = x_{1}x_{2} s$.	We define the 
{\textit{partonic threshold}} as the region of the 
phase space where $\hat{s}\to M^{2}$ ($z\to1$). In 
this regime the dynamics of the process is greatly 
simplified: the partonic center-of-mass energy 
$\sqrt{\hat{s}}$ is just enough to create a 
slepton-pair, and there is no phase space available 
for the emission of hard gluons or soft quarks. 
The cross section is dominated by the terms which 
are singular in the $z\to1$ limit, which correspond 
to the virtual corrections and the real emission of 
soft gluons.
After integrating over the rapidity and retaining 
only the contributions which are singular in the 
threshold region, it is possible to write the 
invariant mass distribution of the process in 
Eq.~(\ref{process}) as
\be
\frac{d \sigma^{\text{thresh}}}{d M^{2}}=
\frac{\pi \alpha_{\text{em}} \beta^{3}_{\tilde{l}}}{3 N_{c} M^{2} s}
\sum_{q}f^{\tilde{l}}_{q}\int^{1}_{\tau}\frac{d z}{z}
C(z,M,m_{\tilde{q}},m_{\tilde{g}},\mu_{f})\ff(\tau/z,\mu_{f})\, ,
\ee
where
\be
f^{\tilde{l}}_{q}=\left[e^{2}_{q} - \frac{e_{q}(g^{q}_{L}+g^{q}_{R}) 
g^{\tilde{l},Z}}{1-m^{2}_{Z}/M^{2}}+
\frac{1}{2}\frac{(g^{q\, 2}_{L}+g^{q\, 2}_{R})g^{{\tilde{l},Z}^{2}}}{(1-m^{2}_{Z}/M^{2})^{2}}\right]\, ,
\ee
and
\be
\ff(y,\mu_{f}) = \int^{1}_{y}\frac{dx}{x}
\left[f_{q/N_{1}}(x,\mu_{f})f_{{\bar{q}/N_{2}}}(y/x,\mu_{f})+(q\leftrightarrow\bar{q})\right]\, ,
\ee
defines the parton luminosity function.
The hard-scattering kernel 
$C(z,M,m_{\tilde{q}},m_{\tilde{g}},\mu_{f})$ depends 
on the invariant mass $M$, on the ratio $z$, as well 
on the masses of squarks and gluinos, 
$m_{\tilde{q}}$, $m_{\tilde{g}}$. By using 
soft-collinear effective theory methods, it is 
possible to prove that, in the $z\to1$ limit, the 
hard-scattering kernel factorize in the following 
way;
\be
C(z,M,m_{\tilde{q}},m_{\tilde{g}},\mu_{f}) = 
H(M,m_{\tilde{q}},m_{\tilde{g}},\mu_{f}) 
S(\sqrt{\hat{s}}(1-z),\mu_{f})\, .
\ee
The hard function $H$ is related to virtual corrections, 
while the soft function $S$ originate from the real emission 
of soft gluons. At \textit{n-}th order in perturbation theory 
the soft function involve plus distributions of the form 
$\alpha^{n}_{s}\left[\ln^{m}(1-z)/(1-z)\right]_{+}$, where $m=0,\ldots,2n-1$.
This needs to be resummed to all orders. The resummation of these 
singular threshold logarithms can be accomplished by solving 
renormalization-group equations for the hard and the soft 
function in the effective theory. To obtain the best possible 
predictions we perform the resummation at NNNLL accuracy and 
then match the result onto NLO calculation.

\section{Phenomenology}
\begin{figure}[t]
\begin{center}
  \includegraphics[width=0.44\textwidth]{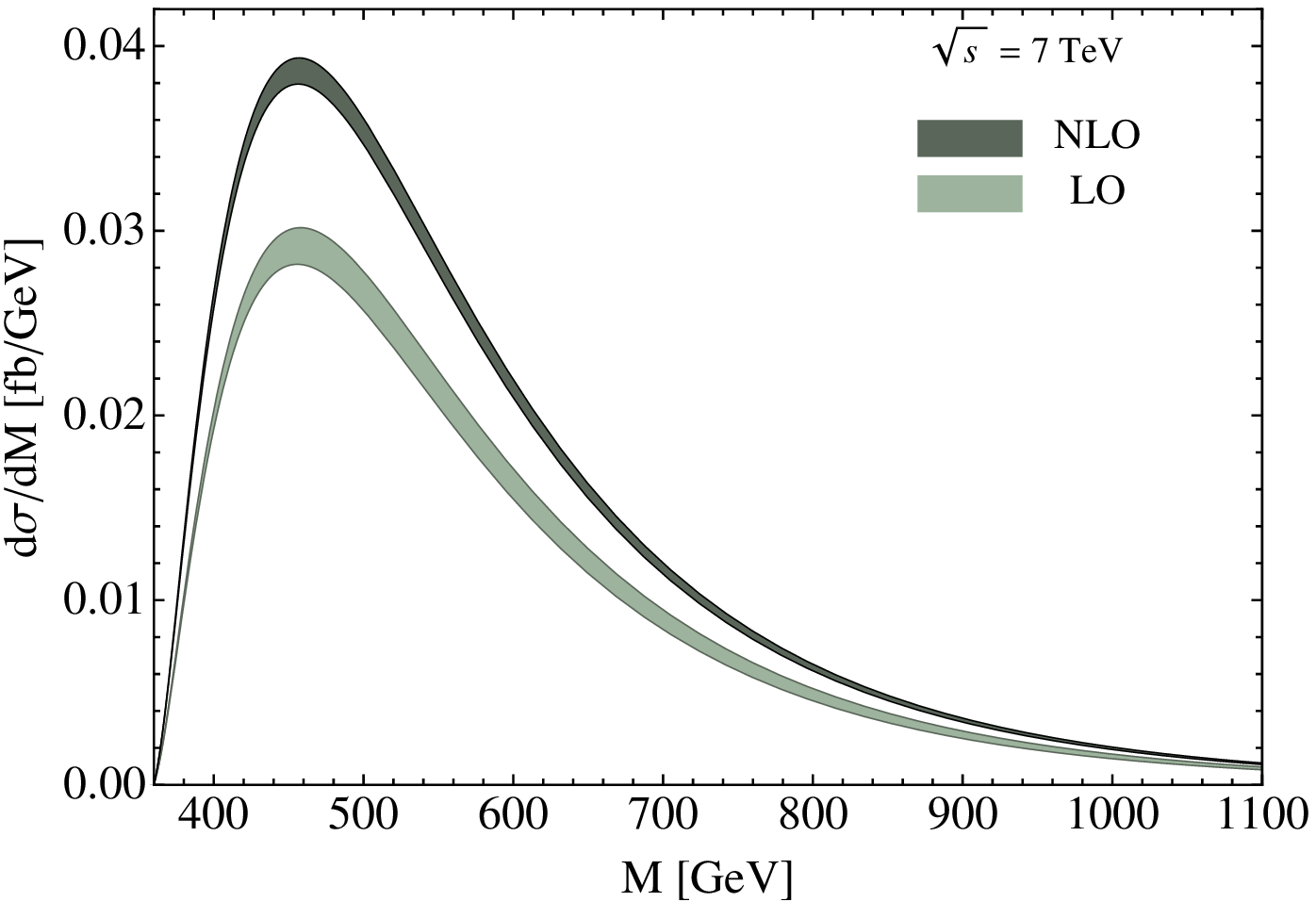}
  \includegraphics[width=0.44\textwidth]{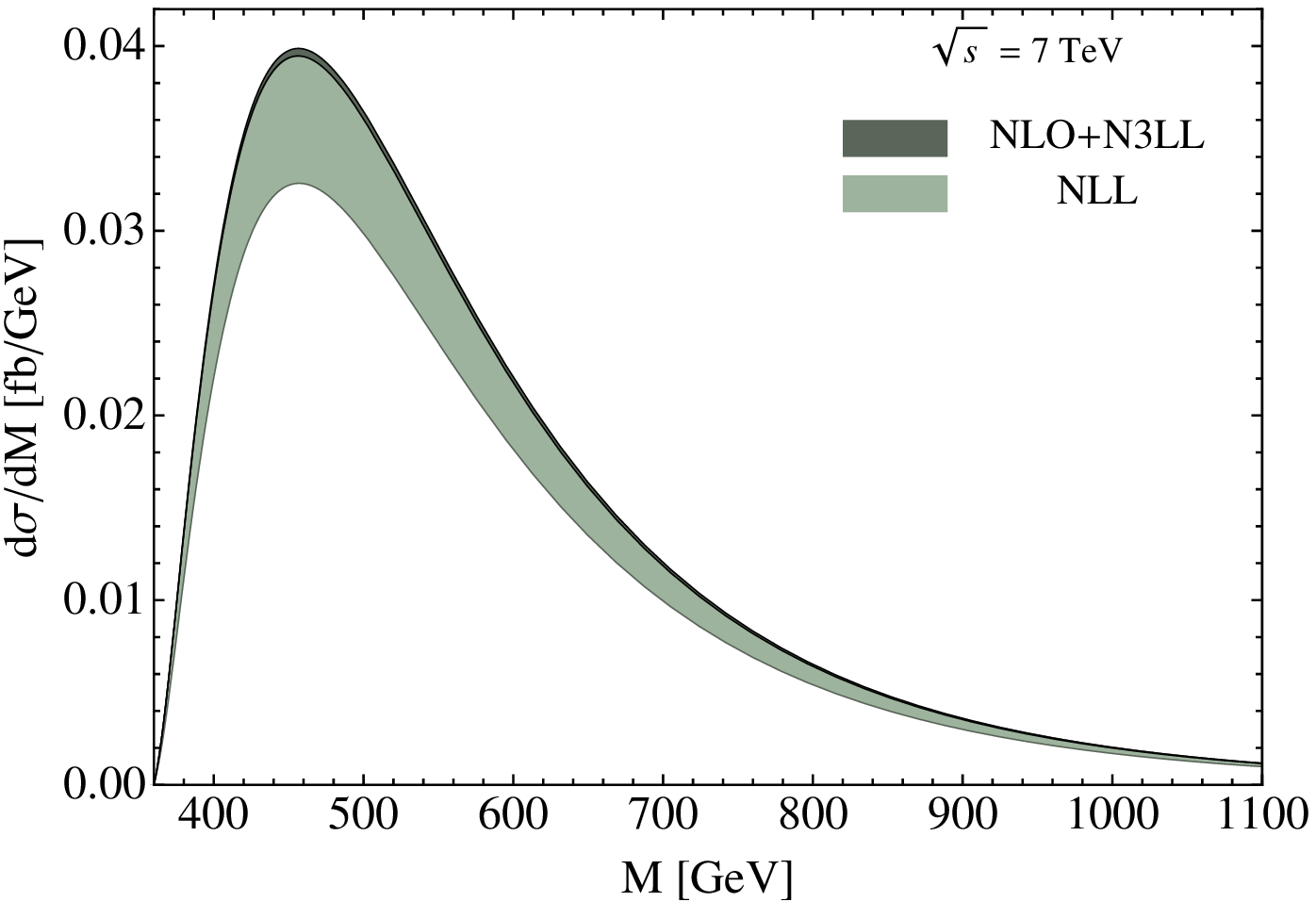}
  \vspace{-0.6cm}
\end{center}
  \caption{Invariant mass distribution for slepton-pair production. These plots are obtained for the following 
  choice of the supersymmetric masses: $m_{\tilde{l}}=180$ GeV, 
  $m_{\tilde{q}}=600$ GeV and $m_{\tilde{g}}=750$ GeV. In the 
  left plots we show the fixed order result at full LO (light 
  band) and NLO (dark band); in the right plot we show the 
  matched result at NLL (light band) and NLO+NNNLL (dark 
  band). The bands refer to the uncertainty associated with 
  the variation of the scales.}
  \label{fig1}
\end{figure}

In this section we show some numerical results for the resummed 
invariant mass distribution and the total cross section for the 
LHC at 7 TeV center-of-mass energy. For a detailed discussion we 
refer to \cite{Mainz1}.
The results for the invariant mass distribution shown in 
Fig.~(\ref{fig1}) have been obtained by using MSTW2008 PDFs 
\cite{Martin:2009iq}. We consider the PDFs at the order which 
is appropriate for the expansion in perturbation theory of the 
corresponding hard-scattering kernel, i.e. we consider LO and 
NLO PDFs for the full LO and NLO distributions, NLO and NNLO 
PDFs for the matched NLL and NLO+NNNLL distributions. As expected, 
the width of the NLO+NNNLL band, obtained varying the hard, soft 
and the factorization scale between 0.5 and 2 times their default 
value is significantly smaller than the band due to scale
uncertainty in the (fixed order) NLO result.

We obtain the total cross section by integrating the invariant 
mass distribution. The numerical results are reported in 
Table~\ref{fig1} (left) with the scales and the PDFs uncertainties. 
In Table~\ref{fig2} (right) is shown the dependence of the total cross 
section from the sleptons masses. The resummation effect for the 
LHC gives an additional contribution of 3 \% on the total 
cross section respect to the NLO, bigger effects, of around 
7 \%, are found at the Tevatron \cite{Mainz1}. 

In conclusion the threshold resummation has a relatively small 
effect on the cross sections, but it reduces significantly the 
scale dependence of the predictions.

\begin{table}[tL]
\begin{center}
  \includegraphics[width=0.396\textwidth]{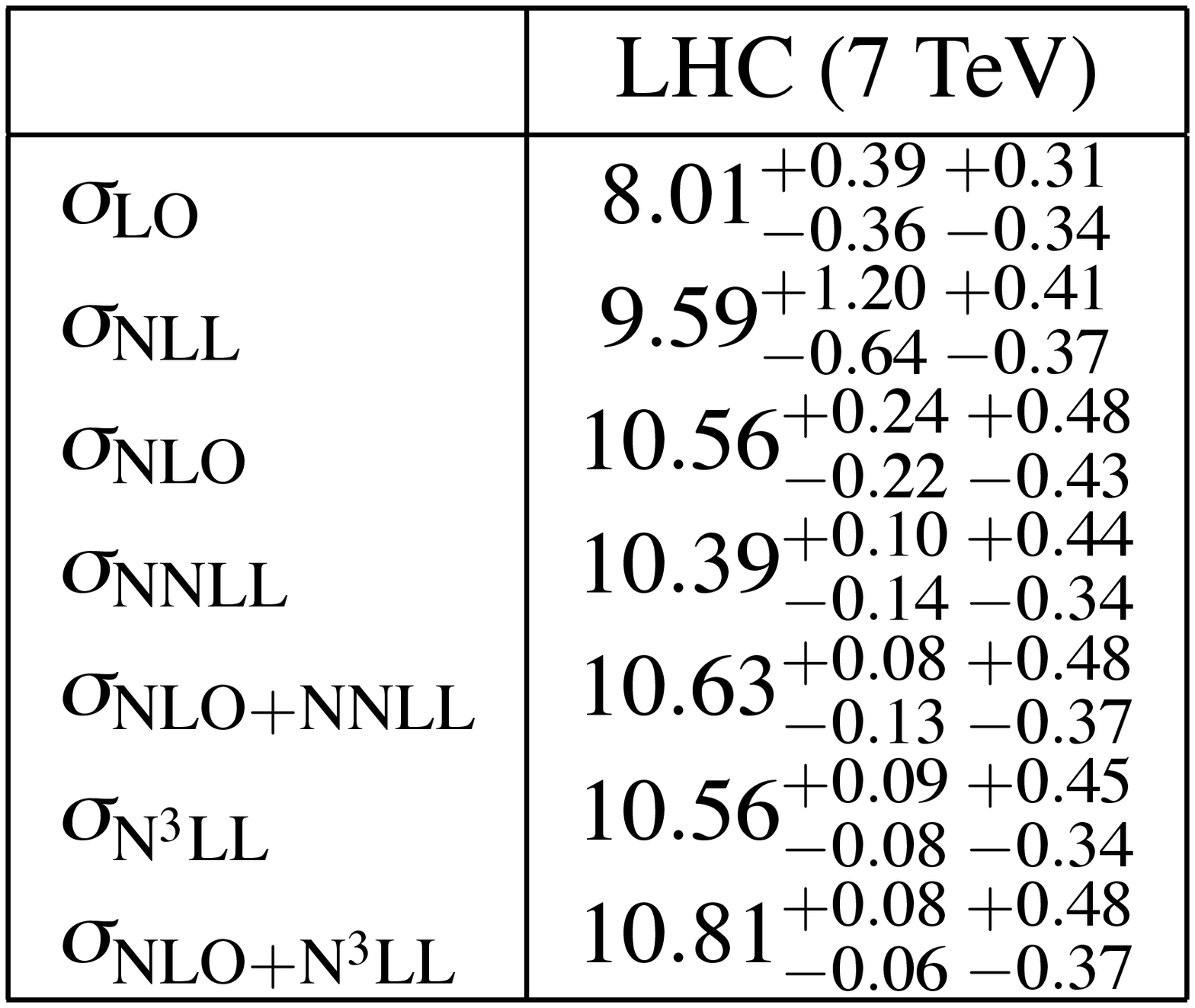} 
  \includegraphics[width=0.54\textwidth]{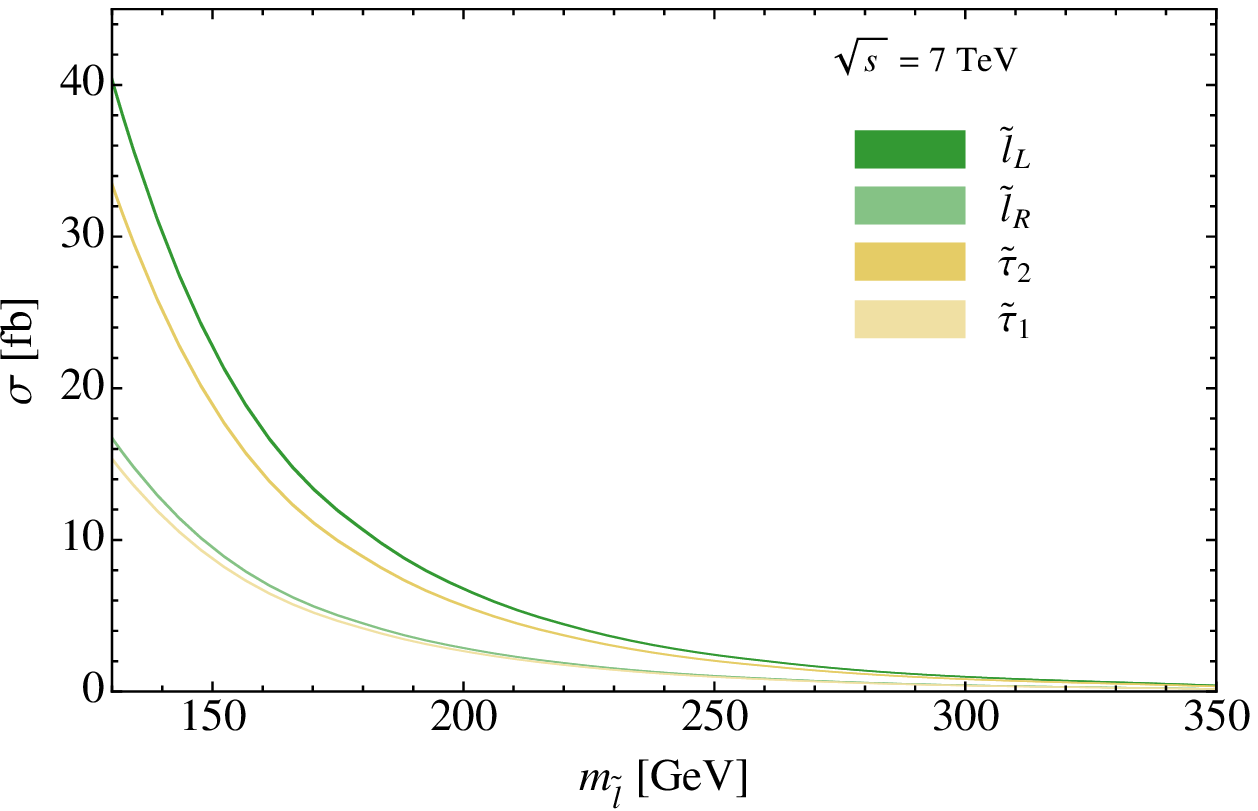}
  \vspace{-0.6cm}
\end{center}
  \caption{Left: total cross sections in fb. The first error refers to the 
  perturbative uncertainties associated with scale variations, the 
  second to PDF uncertainties. The factorization scale $\mu_{f}$ is 
  set equal to the invariant mass $M$. Right: Cross section for slepton-pair production as a function of the slepton mass for
  $m_{\tilde{g}}=750$ GeV and $m_{\tilde{q}}=600$ 
  GeV. This plot refers to our best prediction NLO+NNNLL.}
\label{fig2}
\end{table}


\begin{thebibliography}{99}

%\cite{Beenakker:1999xh}
\bibitem{Beenakker:1999xh}
  W.~Beenakker, M.~Klasen, M.~Kramer, T.~Plehn, M.~Spira, P.~M.~Zerwas,
%  ``The Production of charginos / neutralinos and sleptons at hadron colliders,''
  Phys.\ Rev.\ Lett.\  {\bf 83 } (1999)  3780-3783.
  [hep-ph/9906298].

%\cite{Bozzi:2007qr}
\bibitem{Bozzi:2007qr}
  G.~Bozzi, B.~Fuks, M.~Klasen,
%  ``Threshold Resummation for Slepton-Pair Production at Hadron Colliders,''
  Nucl.\ Phys.\  {\bf B777 } (2007)  157-181.
  [hep-ph/0701202].

\bibitem{Mainz1}
  A.~Broggio, M.~Neubert and L.~Vernazza,
%  ``Threshold resummation for slepton-pair production and Drell-Yan at NNNLL in supersymmetry,''
  in preparation;
  T.~Becher, M.~Neubert, G.~Xu,
  %``Dynamical Threshold Enhancement and Resummation in Drell-Yan Production,''
  JHEP {\bf 0807 } (2008)  030.
  [arXiv:0710.0680 [hep-ph]].

%\cite{Martin:2009iq}
\bibitem{Martin:2009iq}
  A.~D.~Martin, W.~J.~Stirling, R.~S.~Thorne, G.~Watt,
  %``Parton distributions for the LHC,''
  Eur.\ Phys.\ J.\  {\bf C63 } (2009)  189-285.
  [hep-ph/0901.0002].

\end{thebibliography}
\end{document}